\documentclass{article}
\usepackage[utf8]{inputenc}

\usepackage[a4paper, total={6.3in, 8in}]{geometry}
\title{The faint light in groups and clusters of galaxies}
\author{Mireia Montes}

\usepackage{graphicx}

\usepackage[hyphens]{url}
\usepackage{hyperref}
\usepackage{xcolor}
\usepackage{amssymb}
\definecolor{purple}{rgb}{0.26,0.78,0.56}
\hypersetup{
     colorlinks  = true,
     linkcolor   = blue, 
     anchorcolor = BrickRed,
     citecolor   = purple,
     filecolor   = blue,
     menucolor   = blue,
     runcolor    = cyan,
     urlcolor    = blue
}

\begin{document}

\maketitle

The diffuse light that spreads through groups and clusters of galaxies is made of free-floating stars not bound to any galaxy. This is known as the intracluster light (ICL) and holds important clues for understanding the evolution of these large structures. The study of this light has gained traction in the past 20 years thanks to technological and data processing advances that have permitted us to reach unprecedented observational depths. This progress has led to groundbreaking results in the field, such as pinpointing the origin of the ICL and its potential to map dark matter in clusters of galaxies. We now enter an era of deep and wide surveys that promise to uncover the faint Universe as never seen before, adding to our growing understanding of the properties of the ICL and, consequently, of the formation of the largest gravitationally-bound structures in the Universe.
The goal of this review is to summarize the most recent results on ICL, synthesizing the current knowledge in the field and providing a global perspective that may benefit future ICL studies.

\section{General context}
For centuries, the only way humans had to observe the night sky was by naked eye. The faintest stars that the human eye can detect reach 6th magnitude, which yields to around five thousand stars visible to our eyes in some of the darkest skies on Earth. From Aboriginal Australians to Kepler, human perseverance has shown that the most basic instrumentation can still greatly contribute to science.

A revolution occurred in the 17th century: the invention of the telescope. Telescopes have enabled a series of astronomical observations that completely changed our perspective of the Universe. Particularly, the last part of the 20th century saw rapid technological advances in astronomical instrumentation. Bigger and better telescopes paved the way for a new era of discoveries, especially with the advent of space telescopes. At the same time, astronomical detectors became more and more efficient, from the 19th century photographic plates to photometers to, finally, the CCDs that we currently use. Specialized instrumentation allowed us to explore beyond the visible spectrum, including infrared light, radio waves and X-rays.

One of the last remaining frontiers in optical observational astronomy is the low surface brightness Universe ($\mu_V \gtrsim 27$ mag/arcsec$^2$). This is the Universe at the lowest density of stars, largely unseen by past large field surveys like the Sloan Digital Sky Survey (SDSS). The study of this domain promises to deliver transformational insights in our knowledge of the star formation in galaxies in the lowest mass regime, the hierarchical assembly of galaxies and clusters of galaxies and the ultimate nature of dark matter.

\section*{Definition and general properties of the intracluster light} 
Ref. \cite{Zwicky1951} wrote that ``One of the most interesting discoveries made in the course of this investigation is the observation of an extended mass of luminous intergalactic matter of very low surface brightness" in the Coma cluster. For the following seventy years, deep observations of clusters of galaxies have confirmed that first visual discovery: the space between the galaxies in clusters is filled with a faint glow. This faint and extended light is the intracluster light (ICL). For simplicity, in this review I will be referring as ICL to both intragroup and intracluster light. One of the most iconic examples is the ultra-deep image of the Virgo cluster in ref. \cite{Mihos2005}.


Observations during the past 20 years have shown that the ICL is a ubiquitous feature in galaxy clusters \cite{Feldmeier2002, Krick2007, Kluge2020}. But, what is this light? We understand as ICL the light formed by stars that are floating freely inside the cluster's gravitational potential, not bound to any galaxy. It is normally more concentrated around the brightest cluster galaxy (BCG) \cite{Mihos2005, Arnaboldi2012} and shows a variety of morphologies from a featureless smooth component \cite{Iodice2016} to one showing more substructure \cite{Watkins2014, Mihos2017, Kluge2021}.

Observationally, the ICL shows up as an excess of light in the outer parts of BCGs, beyond the extent of the stellar halo of the galaxy. This excess can be described as an extended component over the S\'ersic profile \cite{Sersic1968} describing the light profile of the galaxy. However, this excess or extra component presents a growing ellipticity with radius \cite{Gonzalez2005, Huang2018, Montes2021} and sometimes even a different position angle than the BCG \cite{Kluge2021} meaning that they truly are distinct stellar components. I warn the reader that I refer as ICL to both the stellar halo or envelope (bound to the BCG) and the ICL (bound to the cluster) as they have a common accreted origin and are not distinguishable using imaging alone. 
In cosmological simulations, the stars that belong to this extended component also show a higher velocity dispersion consistent with being bound to the potential of the cluster \cite{Willman2004, Dolag2010, Cui2014, Cooper2015}. In fact, integrated light spectroscopy \cite{Dressler1979, Kelson2002, Edwards2016, Gu2020} and planetary nebulae kinematics \cite{Arnaboldi1996,Longobardi2018} of nearby BCGs, show that the radial velocity dispersion increases with radius reaching the velocity dispersion of the galaxies of the cluster. All this together is suggestive that this extended component is a kinematically distinct component, as confirmed in the nearby Virgo cluster \cite{Longobardi2015}.

Early observations already linked this diffuse light to tidal streams and other signs of interactions in cluster galaxies \cite{Gregg1998, Mihos2005} meaning that the ICL originates from the tidal stripping and merging of galaxies as they interact inside the cluster showing that ICL formation is an ongoing process. Being the byproduct of these processes, over time the ICL forms a fossil record of all the dynamical interactions the system has experienced and offers a holistic view of the history of the cluster. That is, the origin and assembly history of the ICL is central to understanding the global evolution of the cluster galaxy population.

\subsection*{The definition of ICL in observations}

By measuring the amount of light or mass in this component with respect to the total of the cluster, the ICL fraction, we can evaluate the efficiency of the processes that shaped the cluster. Note that the majority of studies use the fraction of light, instead of the mass fraction, in only one photometric filter due to the difficulties in obtaining deep images of clusters in more than one photometric band.

\begin{figure}
\begin{center}
 \includegraphics[width=0.9\textwidth, height = 0.4\textheight]{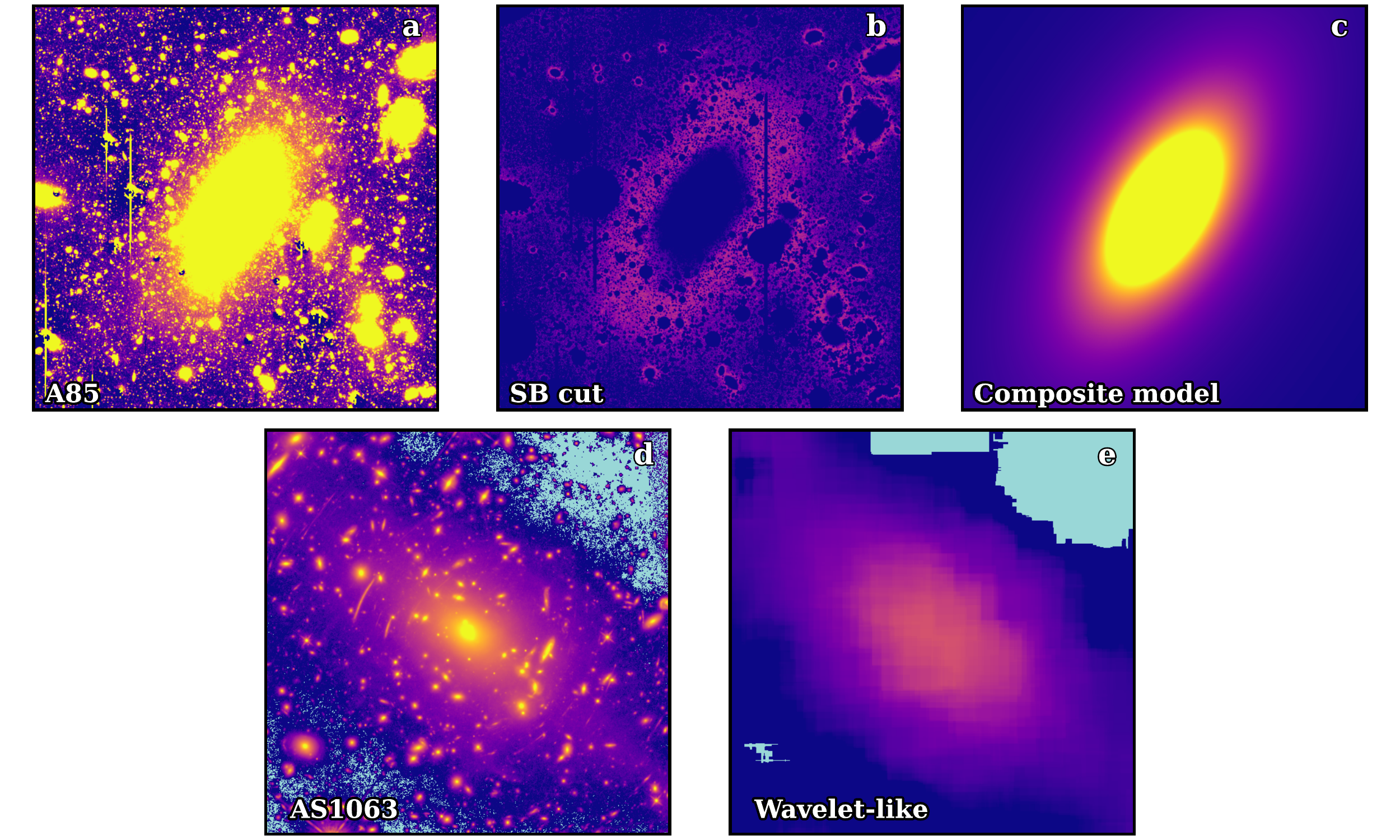} 
 \caption{Example of the different definitions of ICL. \textbf{a}. Image of Abell 85 in the g-band. \textbf{b}. The ICL defined with the SB cut method for Abell 85. \textbf{c}. The BCG+ICL model or composite model for Abell 85. \textbf{d}. Abell S1063 in the F160W band. \textbf{e}. ICL map derived using a 2-D ``wavelet-like'' method for Abell S1063. Adapted from \cite{Montes2021} (Abell 85) and \cite{Morishita2017} (Abell S1063).}
  \label{fig0}
\end{center}
\end{figure}

Although the ICL fraction promises to be an important tool in understanding how clusters evolve, it has a crucial problem: the ambiguous observational definition of the ICL. While in theory one can define the ICL as formed by stars that are not bound to any particular galaxy but to the potential of the cluster \cite{Dolag2010}, in practise this definition is almost impossible to apply in observations and, to a certain extent, in simulations \cite{Canas2020}. Luminous tracers like PNes \cite{Arnaboldi1996} or globular clusters (intracluster globular clusters \cite{Alamo2017, Harris2020}) can trace the kinematics of the stars providing the information to identify the ICL alongside with substructure in the ICL linked to accretion or merger events \cite{Powalka2018}. However, these tracers can only be studied in the nearest clusters (as far as the Coma Cluster) \cite{Madrid2018}, but for most clusters this information is not available.

Deep images of clusters of galaxies show that the transition between BCG and ICL happens smoothly with no clear break point \cite{Gonzalez2005, Seigar2007, Iodice2016, Montes2021}. This is a natural consequence of the fact that BCGs are made up of as much as $70\%$ of accreted, ex-situ material \cite{Pillepich2018}; the same material that forms the ICL. Therefore, without kinematical information available, observers had to come up with a way of separating both components to be able to study the ICL and its evolution separately. 
 
There are three different methods that are mainly used in the literature. The first, and easiest from an observational point of view, is to use a surface brightness cut (SB cut) to separate between the galaxies and the ICL. This method is based on assuming that all the light below a certain surface brightness is ICL. Simulations have proven that this method is effective in separating the BCG and the diffuse light \cite{Rudick2011, Cui2014} at a surface brightness fainter than $\mu_V > 26.5$ mag/arcsec$^2$. However, using this definition in observations is a more complicated issue as different different image depths lead to different amounts of ICL \cite{Burke2015, MT18, Tang2018}. In addition, each observer uses a different band and SB limit to define the ICL in their observations, making direct comparison difficult. The SB cut method is shown in the top middle panel of Fig. \ref{fig0} for the g-band of Abell 85 (adapted from \cite{Montes2021}). 

Another widely use method is to assume that BCG and ICL can be described by adding together multiple models (composite model) such as S\'ersic or exponential models. Although this has the advantage of separating both components, it can be heavily degenerate. For example, \cite{Janowiecki2010} showed that galaxies in Virgo can be described by either a single or double S\'ersic models resulting in very different ICL fractions. This method also fails to account for all the diffuse light associated with satellite galaxies, i.e., that is not concentrated around the BCG, asymmetries in the light distribution and substructure like tidal streams. As an example, a model of the BCG+ICL of Abell 85 (g-band) is shown in the top right panel of Fig. \ref{fig0}. 

Recently, there has been a push for more complex algorithms aiming at separating galaxies and diffuse light in two dimensions. This has been done using 2D fitting algorithms like GALFIT to most of the galaxies in the cluster \cite{Giallongo2014, Morishita2017} or more sophisticated ``wavelet-like'' decomposition techniques \cite{daRocha2005, Jimenez-Teja2016, Ellien2021}. This approach might solve most of the shortcomings of the previous two methods as it separates galaxies and ICL for the whole image. An example of this approach can be found in the bottom right panel in Fig. \ref{fig0} for the F160W band of Abell S1063 (priv. comm. \cite{Morishita2017}).

The examples in Fig. \ref{fig0} show a clear variance in the measured ICL fraction depending on the method used. For example, for Abell 85 (top panel) when using the SB cut method ($\mu_g > 26.5$ mag/arcsec$^2$) the ICL fraction measured is $6.5\pm0.7$\% while the composite model gives a fraction of $11\pm1$\% \cite{Montes2021}. As mentioned before, this is expected because the composite models method extrapolates the diffuse light in the line of sight of the BCG which results in adding more ICL that the SB cut method cannot account for. In the same way, for the ``wavelet-like'' method (bottom row), the ICL fraction in Abell S1063 is $23\pm3$\% (ICL mass fraction \cite{Morishita2017}) while \cite{MT18}, using the SB cut method ($\mu_V>26$ mag/arcsec$^2$), found an ICL fraction of $3.24\pm0.03$\%. While these two measurements are not directly comparable, it is clear that the ``wavelet-like'' method provides a more complete picture of the ICL. 

The previous comparison illustrates the uncertainty in measuring the ICL fraction in observations. Simulations have already shown that ICL fractions can vary up to a factor of two depending on the method used to define the ICL \cite{Rudick2011}. In observations, ref. \cite{Kluge2021} found a scatter ranging from $34\%$ to $71\%$ in their ICL fractions when using $4$ different methods in their sample of $170$ clusters of galaxies. Additionally, ICL fractions are affected by observational parameters such as cosmological dimming (or how far we can explore the ICL) or the spatial resolution of the images \cite{Tang2018}. 

This highlights how important it is to use a consistent metric to measure ICL fractions, especially if we want to compare between clusters and derive correlations with mass and redshift. If not, any conclusion will be meaningless. Bringing together observers and simulators to find an adequate method to measure ICL fractions is more crucial now than ever. One thing that the community should consider is that given that most the stellar mass of the BCG is accreted, it might not make sense to separate both components but treat them as one to avoid ambiguity (as suggested in ref. \cite{Gonzalez2007}).

\subsubsection*{Stellar populations}

A useful tool to characterize the ICL is the study of its stellar populations, as they reflect the properties of the galaxies from which the ICL accreted its stars. Knowing the stellar populations of the ICL in clusters allows us to infer the mechanisms at play in the formation of this component, and therefore how (and when) the assembly history of these clusters occurred. 

Simulations have provided different mechanisms that can be responsible for forming the ICL: total disruption of low mass satellites \cite{Purcell2007}, tidal stripping of massive satellites \cite{Rudick2009, Contini2014}, stars ejected into the intracluster medium after a merger \cite{Willman2004, Murante2007, Conroy2007} and accretion of ICL from groups (pre-processing, \cite{Mihos2004, Rudick2006, Contini2014}). 

Given the characteristics of this light, faint ($\mu_V > 26.5$ mag/arcsec$^2$, \cite{Rudick2011}) and extended (several hundreds of kpc), for most clusters the only information available is via broadband photometry. This means that most of our current knowledge of the stellar populations of the ICL is from the dominant mechanism of formation, rather than of the, likely, most complex picture. Fortunately, each of the aforementioned mechanisms will leave a distinct imprint in the stellar populations of the ICL. For example, the properties of the ICL, color and metallicity, will be very different if it formed mostly through tidal stripping of massive satellites or the total disruption of dwarf galaxies or if the stars are ejected after major mergers. A more in-depth review on the mechanisms of the formation of the ICL from the point of view of simulations can be found in ref. \cite{Contini2021review}. 

Studies have found clear radial negative gradients in color (bluer with radius \cite{Krick2007, Rudick2009, Melnick2012, DeMaio2015, Iodice2017}) indicating radial gradients in the stellar populations of the ICL. The colors of the ICL are found to be similar to the average colors of the satellite galaxies in the clusters. 
In general, these gradients can be explained as radial changes in metallicity \cite{MT14, DeMaio2015, Gu2020} and, in some cases, age \cite{Morishita2017, MT18}. Both colors and metallicities indicate that the progenitors of the ICL are the stars in the outskirts of galaxies of masses around $5\times10^{10}$ M$_\odot$ \cite{MT18, Montes2021}. The scenario where the total disruption of dwarf galaxies is the main contributor to the ICL is ruled out as the number of galaxies required to explain the total luminosity of the ICL will dramatically change the luminosity function of clusters \cite{DeMaio2018}. On the other hand, there is evidence of a few clusters showing flat gradients in color indicating ICL formation through stars expelled into the intracluster medium in a major merger \cite{Krick2007, DeMaio2015, DeMaio2018}. This suggests that the mechanism of formation of the ICL might depend on the specific assembly history of each cluster. For example, the age and metallicity profiles of Abell 370 remain flat down to $\sim50$ kpc likely due to the interaction between the two BCGs of the cluster \cite{MT18}. 

Regarding the age of the ICL, for intermediate redshift clusters it is found to be 2-6 Gyr younger than the BCG \cite{Toledo2011, Adami2016, MT18} pointing to the stripping of stars from galaxies that stopped forming stars when they entered the cluster \cite{Morishita2017, MT18}. Additionally, ref. \cite{Jimenez-Teja2018} found that merging clusters present a higher ICL fraction at blue wavelengths (approximately rest-frame B-band). This is indicative of a younger stellar population in the ICL in merging clusters and, therefore, that the stars that form the ICL might have been stripped from recently quenched galaxies during the merging process. Given that the ages of the ICL in nearby clusters are, for the most part, old \cite{Williams2007, Coccato2010}, it seems that it evolves passively without adding a significant fraction of younger stars.

In the case of groups of galaxies, observations have also found negative color gradients of their diffuse light \cite{Hartke2017, DeMaio2018, Spavone2018, Ragusa2021}. In this case, the color of the ICL is similar to the mean color of the massive galaxies of the group, pointing to the interaction between these galaxies as the source of ICL \cite{daRocha2005, Spavone2018}. 

There are a handful of systems that, due to their proximity, have been studied in more detail and show a more complex picture of ICL formation. One of those systems is the Virgo cluster. Ref. \cite{Williams2007} measured RGB stars in a Hubble Space Telescope (HST) pointing of the cluster finding that that most of the ICL ($\sim 75\%$) comes from old, $t>10$ Gyr, low-metallicity stars (median [M/H]$\sim -1.3$) but $\sim25\%$ have higher metallicities and intermediate ages ([M/H]$> -0.5$, t$<10$ Gyr). This indicates that, in that particular field, dwarf galaxies are a significant source of ICL stars. This, of course, might not be representative of the ICL in Virgo as the ICL stars in this cluster are not well mixed \cite{Arnaboldi2004, Mihos2005}, indicating that different processes might act in different regions of the cluster.  Using the same technique, ref. \cite{Smercina2020} found that while the current ICL in the M81 group is metal-poor ([Fe/H] $\sim -1.2$), the ongoing accretion of the other members onto M81 will form a massive, more metallic halo ([Fe/H] $= -0.9$). This suggests that the processes that contribute to ICL formation will depend on the evolutionary state of the group or cluster. 
The level of information and detail that these nearby systems provide to our knowledge of the formation of the ICL is invaluable \cite{Iodice2017} and more detailed studies of these systems are needed.

Recent simulations predict that the bulk of the ICL mass at recent times comes mainly from the tidal stripping of massive satellites ($10<\log (M/M_{\odot})<11$) \cite{Contini2014, Contini2019}. Disk-like massive satellites make the largest contribution to building the ICL through a large number of small stripping events \cite{Contini2018}, in agreement with observations where the metallicity of the ICL is that of the outer parts of massive satellites \cite{MT18}. These simulations have been also able to reproduce the observed metallicity gradients \cite{Cui2014, Contini2014, Contini2019}. Ref. \cite{Cui2014} suggested that the average metallicity of the ICL increases mildly with halo mass, while ref. \cite{Contini2019} found no correlation. In addition, ref. \cite{Contini2019} claims that lower mass satellites ($9<\log (M/M_{\odot})<10$) should be the main contributor to the formation of the ICL at earlier times ($z>0.5$). Observations, so far, seem to favor the lack of correlation with halo mass in ref. \cite{Contini2019}, but not the contribution of lower mass satellites at $z>0.5$ \cite{MT18, DeMaio2018}. However, this comparison between observations and simulations is only based on a handful of observed clusters at intermediate redshifts. Larger samples are needed in order to explore this issue in detail. 

\subsubsection*{The correlation of the ICL fraction with redshift and mass}
Depending on the dominant processes and the epoch at which they occur, the ICL fraction can correlate with global cluster properties such as mass and redshift. For example, a trend of the ICL fraction with redshift will tell us about the timescales of the mechanisms involved in stripping stars from galaxies.

\begin{figure}
\begin{center}
 \includegraphics[width=0.6\textwidth, height = 0.4\textheight]{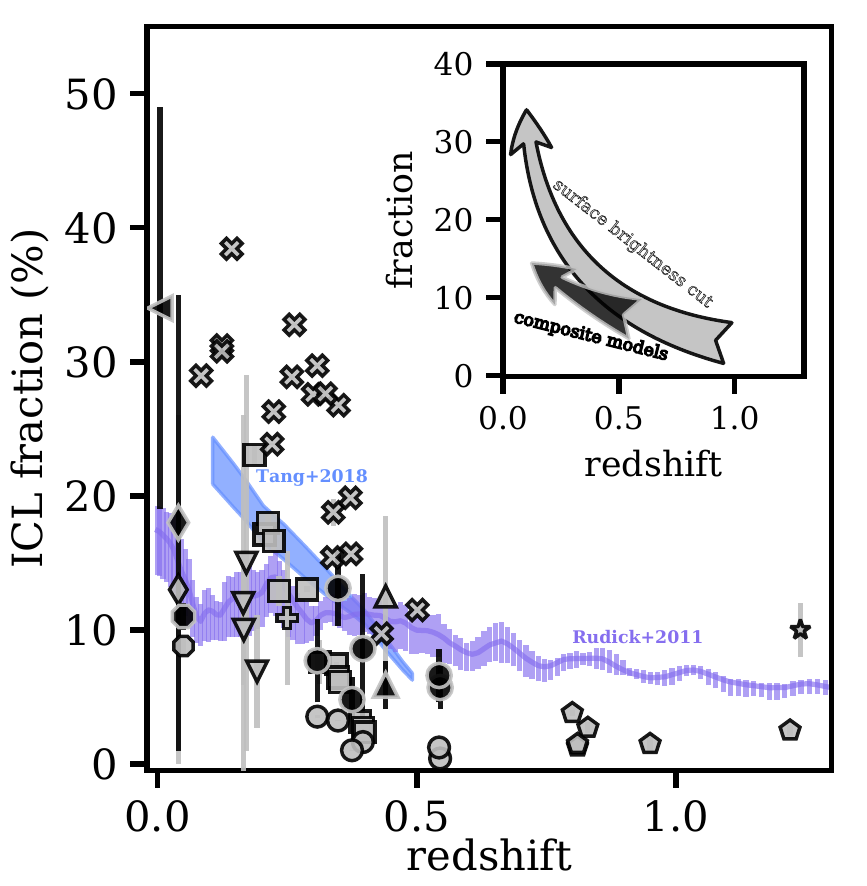} 
 \caption{ICL fraction with redshift for clusters of galaxies. The black symbols indicate those ICL fractions derived using composite models to fit BCG and ICL while the grey symbols indicate those derived using a SB cut. The different markers, and their corresponding errors, indicate the studies were the fractions were taken from. Squares: \cite{Burke2015}, pentagons: \cite{Burke2012}, circles: \cite{MT18}, x: \cite{Furnell2021}, down-pointing triangles: \cite{Feldmeier2004}, octagons: \cite{Montes2021}, diamonds: \cite{Kluge2021}, cross: \cite{Zibetti2005}, star: \cite{Ko2018}, triangles: \cite{Presotto2014}, left-pointing triangle: \cite{Spavone2020}. The purple line and errors correspond to the the simulations of ref. \cite{Rudick2011} for $\mu_V>26$ mag/arcsec$^2$. The blue polygon represents the predictions in ref. \cite{Tang2018}. }
  \label{fig1}
\end{center}
\end{figure}

In Fig. \ref{fig1}, I show how the ICL fraction correlates with redshift for clusters ($\gtrsim 10^{14}$ M$_\odot$, to minimize biases due to possible mass dependencies of the ICL fraction) from measured values in the literature (listed in the caption). The grey markers are the ICL fractions measured using a SB cut while the black markers show those measured using the composite model method. Note that this comparison has to be taken with a grain of salt as the different studies are using different SB cuts, bands and depths. However, there are some common trends that show up. When using a SB cut to measure the ICL, there is a trend of increasing ICL fraction with decreasing redshift. While at $z>0.6$ the ICL fraction remains roughly constant around a few percent, at $z\sim0.6$ the ICL begins to buildup to reach values of $\sim30\%$ at $z=0.1$. However, the composite model method does not show such an steep increase at $z<0.6$. In fact, the latter ICL fractions show little to no evolution. Nonetheless, the composite model method has only been applied to a few clusters and none at $z>0.6$ and therefore it is not possible to compare both methods fairly. The top right panel in Fig. \ref{fig1} shows a schematic diagram with both trends. Note that a lack of correlation with redshift is also present when using ``wavelet-like'' algorithms \cite{Guennou2012, Jimenez-Teja2018}.

From the point of view of simulations, there is a general agreement that the ICL fraction increases with time at $z<1$, from almost no ICL at $z=1$ to around $15-20\%$ of the total light of the cluster at $z = 0$. I included the ICL fraction predictions of ref. \cite{Rudick2011} (purple line) and ref. \cite{Tang2018} (blue polygon) in Fig. \ref{fig1}. When using a SB cut to replicate observations, the simulated ICL seems to be building up more gradually since $z=1$ than seen in observations (see Fig. \ref{fig1}, \cite{Rudick2011, Tang2018, Furnell2021}). Ref. \cite{MT18} argued that this might be caused by the photometric band chosen to derive the ICL fractions (V in simulations vs. B in observations \cite{Rudick2011, Burke2015, Furnell2021}). Bluer optical bands fade rapidly with age and using the same SB cut over a large redshift range could mean including light from the BCG to the ICL, as the isophote of a given surface brightness gets closer to the centre as cosmic time progresses, and stellar populations get older and fainter. In fact, this effect might be causing the disagreement between the observed trends with redshift when using different ICL definitions. A way of overcoming this issue could be using either redder photometric bands or the ICL mass fraction, more robust against drastic brightness changes in the stellar populations due to age evolution.

Groups are an intermediate stage between clusters and the field where environment has started to affect the properties of member galaxies. Simulations suggest that at $z = 0$, almost half of the galaxies in clusters have joined their host cluster from group halos \cite{Mcgee2009, Bahe2013}. This suggests that the formation of the diffuse light in groups of galaxies could contribute significantly to the diffuse light in clusters \cite{Mihos2017, Iodice2017}, a process dubbed as pre-processing like its galaxy analogue. Ref. \cite{Contini2014} showed that up to $30\%$ of the ICL in simulations comes from pre-processing in groups and clusters. 

\begin{figure}
\begin{center}
 \includegraphics[width=0.8\textwidth, height = 0.3\textheight ]{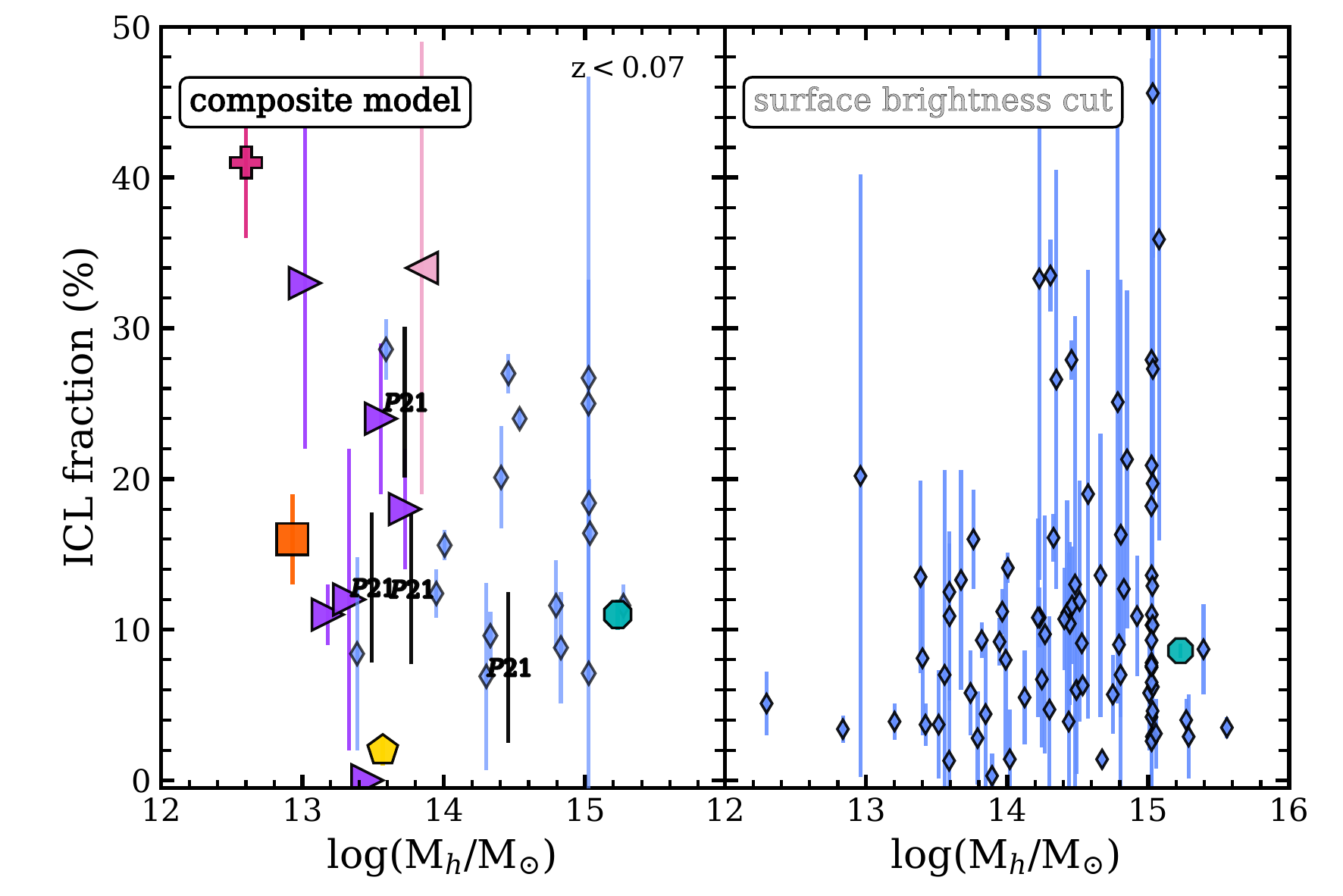} 
 \caption{ICL fraction with halo mass for groups and clusters at $z<0.07$. Left panel shows those fractions derived using composite models to fit and separate BCG and ICL, while the right panel shows the fractions derived using a SB cut. The different markers, and their corresponding errors, indicate the studies were the fractions were taken from. Blue diamonds: \cite{Kluge2021}, teal octagon: \cite{Montes2021}, light pink left-pointing triangle: \cite{Spavone2020}, purple right-pointing triangles: \cite{daRocha2005, daRocha2008}, orange square: \cite{Ragusa2021}, pink cross: \cite{Iodice2020},  yellow pentagon: \cite{Spavone2018}, `P21': \cite{Poliakov2021}}
  \label{fig2}
\end{center}
\end{figure}

Exploring how the ICL fraction correlates with the total mass of the system can give us insights into this mechanism. Fig. \ref{fig2} shows the ICL fraction as a function of the total mass of the cluster for groups and clusters at $z<0.07$. This redshift limit has been chosen in order to minimize any biases due to any possible correlation of the ICL fraction with redshift. The left panel shows the fraction derived for the composite model and the ``wavelet-like'' method while the right panel shows the fractions derived using a SB cut. 

Fig. \ref{fig2} shows no correlation between the ICL fraction and the total cluster mass (see also ref. \cite{Sampaio-Santos2021}). This lack of correlation could mean that the mechanism of formation of the ICL in groups and clusters is similarly efficient. That is, both groups and clusters produce the same quantity of ICL per unit of total mass. Note that there is an increasing dispersion in the ICL fractions with mass for the SB cut method, likely due to a higher number of clusters observed at higher masses, while the dispersion in the composite model appears to be similar at all masses. Perhaps, it will be more informative to see how the ICL fraction correlates with the total stellar mass of the system, the fuel of the ICL, to explore how efficient each system is in forming ICL.

On the theoretical side, simulations do not agree on the expected correlation of the ICL with mass. Some simulations show no trend in the group-cluster mass range \cite{Dolag2010, Contini2018} while others show conflicting trends with mass (increasing \cite{Lin2004, Murante2004, Purcell2007} or decreasing \cite
{Cui2014}). This disparity in the trends might be caused by how differently the ICL forms in these different simulations, whether it is the actual formation mechanism in SAM-type simulations or the resolution and/or underlying physics in numerical simulations.

There is a hint in observations of a correlation between the ICL fraction and the concentration of galaxies in groups so that compact groups show a higher ICL fraction \cite{Aguerri2006, daRocha2008}. However, some nearby loose groups show a significant ICL fraction \cite{Spavone2017, Spavone2018}. Note that the higher ICL fraction in groups might be caused by compact groups being over-represented, as they are easier to observe. There is evidence that the most relaxed, evolved groups and clusters have higher ICL fractions compared to less evolved ones \cite{Aguerri2006, daRocha2008, MT18, Poliakov2021}. This evidence is consistent with expectations from simulations \cite{Sommer-Larsen2006, Rudick2006, Contini2014}. The ICL will grow slowly as substructure (e.g., tidal streams) becomes well-mixed in the cluster potential. If a small system enters the cluster, the quantity of ICL increases rapidly as the new infalling galaxies are exposed to their new and denser cluster environment and also, probably, by accreting pre-processed ICL \cite{Rudick2011}. Therefore, the dynamical state of a group or cluster might be a better tracer of the growth of the ICL fraction than redshift.

\subsubsection*{The link between BCG and ICL}

The formation and evolution of BCGs have been predicted to be rather different than other galaxies: their innermost regions formed the majority of their stars at high redshift and on short timescales \cite{Thomas2005} whereas their outer parts are likely assembled as a consequence of multiple minor merging \cite{Trujillo2011}. As the ICL tends to be concentrated around the BCG \cite{Mihos2005}, it implies that the growth of both components, BCG and ICL, are linked. This scenario is also supported by simulations where up to $\sim70\%$ of the stellar mass of the BCG is accreted indicating that BCG and ICL are formed, mostly, in a similar way \cite{Pillepich2018}. 

Observations show that most BCGs were already in place by $z\sim1$ \cite{Collins2009}. When trying to infer their mass growth with time, observational studies disagree. Since $z=1$, the inferred rate ranges from no growth to a factor of 2 depending on the study \cite{Collins2009, Lidman2012, Zhang2016}. Nonetheless, this is far from the rates expected in simulations (a factor of 4 \cite{deLucia2007}). A way to bring observations and simulations together is if simulations include formation of the ICL. For example, ref. \cite{Conroy2007} replicated observations of nearby clusters by placing a significant fraction of the accreted mass into the ICL component ($\sim50\%$ \cite{Laporte2013, Contini2018}). Fig. 1 in ref. \cite{Montes2019} provides a visual summary of this issue.

\begin{figure}
\begin{center}
 \includegraphics[width=0.75\textwidth, height = 0.3\textheight]{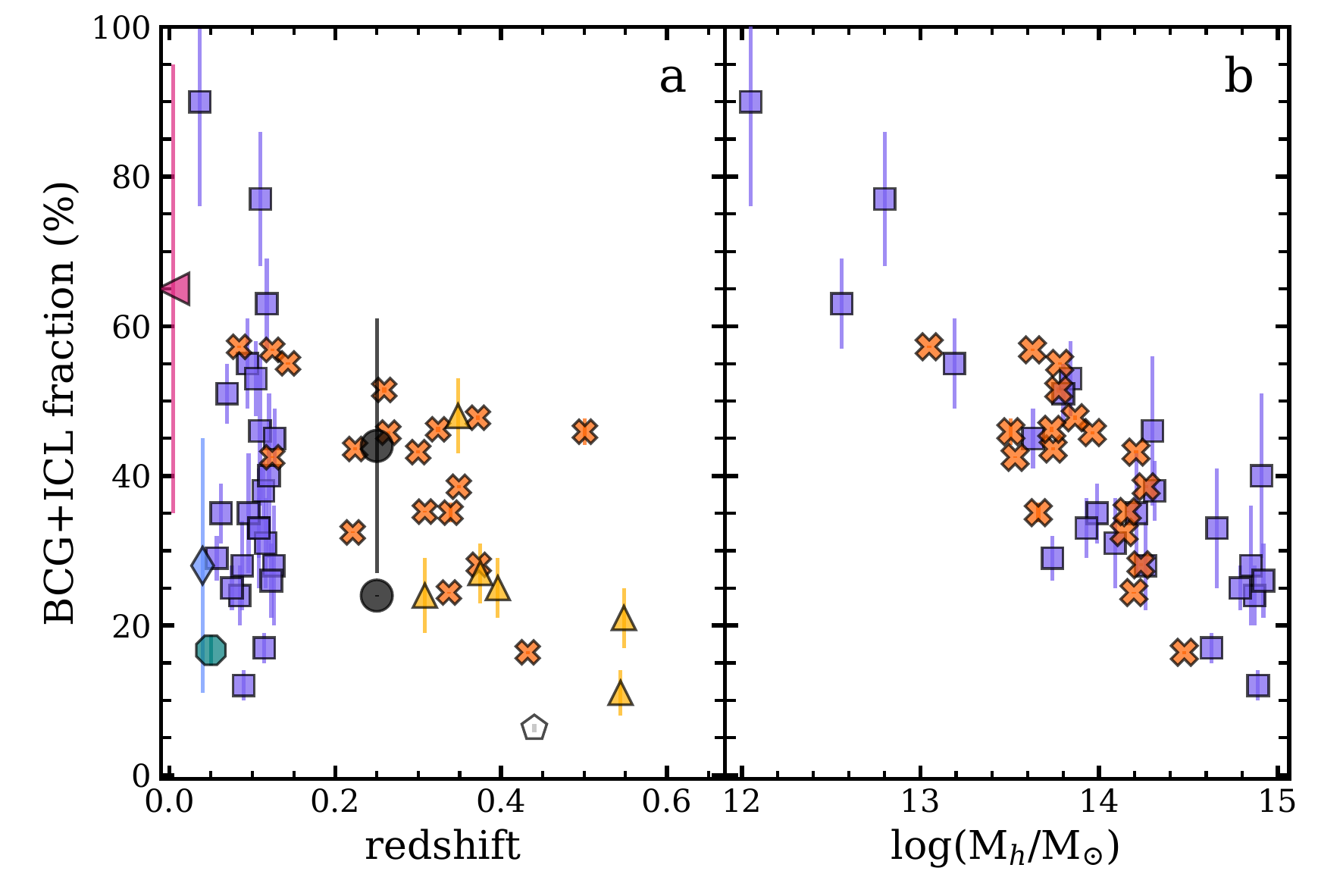} 
 \caption{Correlations of the BCG+ICL fraction. a. with redshift. b. with mass. The different markers and errors indicate the different works the fraction has taken. Orange crosses: \cite{Furnell2021}, purple squares: \cite{Gonzalez2007}, teal octagon: \cite{Montes2021}, black circles: \cite{Zhang2019} and \cite{Sampaio-Santos2021}, yellow triangles: \cite{Morishita2017}, blue diamond: \cite{Kluge2021}, pink left-pointing triangle: \cite{Spavone2020}, white pentagon: \cite{Presotto2014}}
  \label{fig3}
\end{center}
\end{figure}

As mentioned before, given that studying both components together might alleviate the ambiguity in defining the ICL, exploring how the BCG+ICL fraction (the fraction of light in the BCG+ICL with respect to the total light of the cluster) correlates with redshift can give us clues about their common evolution. Fig. \ref{fig3} shows the correlation between the BCG+ICL fraction with redshift (left panel). There seems to be a correlation between BCG + ICL fraction and redshift. However, when exploring the correlation with the total mass of the cluster (right panel), the correlation with redshift is driven by groups at lower redshift rather than a true evolution of this fraction with redshift. 

As seen before, there is a strong negative correlation of the BCG+ICL fraction with the mass of the halo \cite{Gonzalez2005, Gonzalez2007, Burke2015, Furnell2021}. This correlation is telling us that with higher masses, there is more mass located in the satellites of the cluster than in the BCG+ICL component \cite{Gao2004, Gonzalez2013}. In agreement with the lack of correlation with redshift seen in the left panel, there is no significant evolution in this relation between the intermediate-redshift (orange crosses \cite{Furnell2021}) and the low-redshift sample (purple squares \cite{Gonzalez2007}).

\begin{figure}
\begin{center}
 \includegraphics[width=0.45\textwidth]{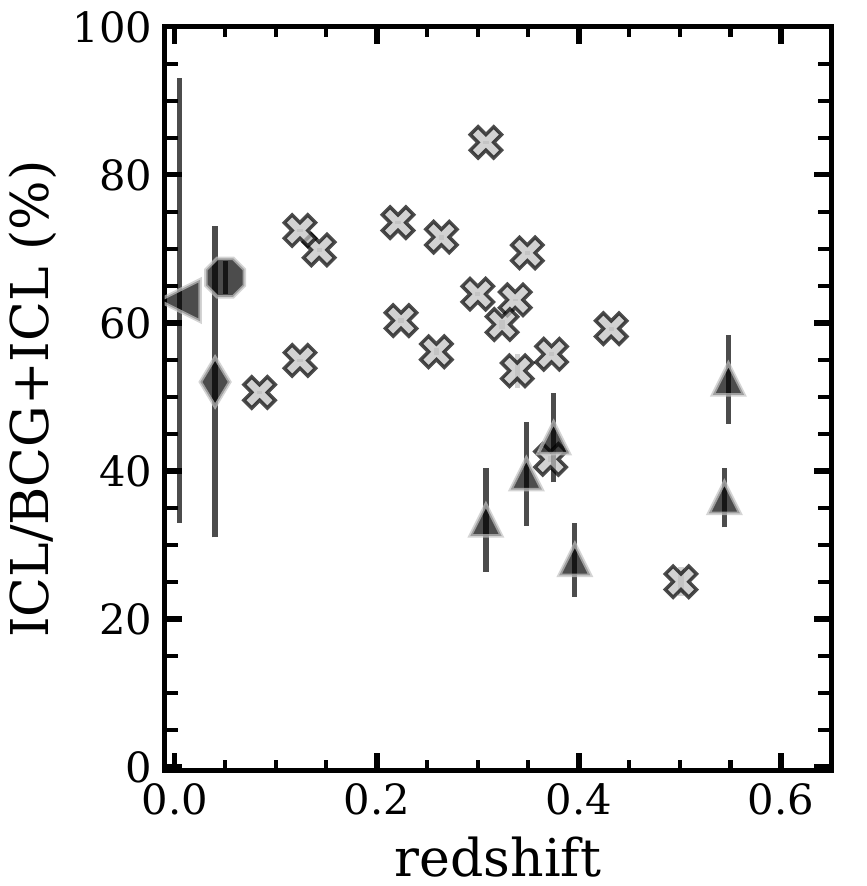} 
 \caption{ICL/BCG+ICL as a function of redshift. The black symbols indicate those ICL fractions derived using a composite model to fit BCG and ICL while the grey symbols indicate those derived using a SB cut. The different markers, and their corresponding errors, indicate the studies were the fractions were taken from. Crosses: \cite{Furnell2021}, octagon: \cite{Montes2021}, triangles: \cite{Morishita2017}, diamond: \cite{Kluge2021}, left-pointing triangle: \cite{Spavone2020}.}
  \label{fig4}
\end{center}
\end{figure}

It is also worth exploring how the fraction of ICL with respect to the total BCG+ICL light correlates with redshift (Fig. \ref{fig4}). Contrarily to what it is expected if the ICL grows at later times but the BCG does not, this fraction remains fairly constant across redshift. Ref. \cite{DeMaio2020} found that the stellar mass of BCG+ICL between 10 to 100 kpc increases by a factor of 2 between $z = 1.55$ to $z= 0.4$, while no significant change is seen from $z= 0.4$ to $z = 0$.  Current observations typically only ``see'' the ICL out to a $100-200$ kpc from the center of the BCG. Ref. \cite{Zhang2019}, stacking $\sim300$ DES clusters at $z\sim 0.25$, observed the ICL up to $\sim1$ Mpc finding that the BCG+ICL profile could be described with $3$ models, with the outermost starting to dominate at $\sim 400$ kpc (see also \cite{Sampaio-Santos2021, Chen2021}). Moreover, ref. \cite{Gonzalez2021} found that over $90\%$ of the total BCG+ICL luminosity lies outside of the central $100$ kpc. If the ICL at $z<0.4$ builds outside $\sim100$ kpc \cite{DeMaio2018}, it might be that we are only observing the ICL already accreted at $z\sim 0.5$ and even deeper observations are needed to assess the true growth of the BCG+ICL in the last few Gyr. This agrees with the semi-analytical simulations of ref. \cite{Contini2018} (see also \cite{Murante2007, Rudick2011}) that show that the ICL grows at later times ($z<0.5$) once the BCG has already (mostly) formed.

\subsection*{The assembly of galaxy clusters and tracing dark matter}

As the ICL forms through the interactions of satellites in the cluster, it encodes crucial information about the global cluster properties and the evolution of galaxy clusters. Ref.\cite{Kluge2021} found positive correlations between BCG+ICL brightness and several properties of the cluster like cluster mass \cite{Zhang2019, Sampaio-Santos2021}, and integrated satellite brightness \cite{Canas2020}, confirming that ICL, or BCG+ICL, is coupled with cluster growth. Recently, ref. \cite{Zhang2019} and ref. \cite{Sampaio-Santos2021} found that the ICL is self-similar. That is, after scaling with cluster radius, the ICL profiles of clusters in different richness ranges are indistinguishable. This is the case for the dark matter, the hot gas, and the radial distribution of member galaxies and connects those different properties with the total mass of the cluster. 

A significant anisotropy in the orientation of the orbits of the progenitors of the ICL will produce an elongation in the ICL distribution. This elongation will increase with radius up to the value of the original distribution, i.e., the cluster distribution \cite{Shin2018}, naturally explaining the increasing ellipticity with radius seen for the ICL \cite{Gonzalez2005, Huang2018, Montes2021}.

In the last few years, the ICL has been shown to be a particularly effective tool to explore dark matter inside clusters of galaxies. The physical scales of the diffuse light, several hundreds of kpc, are similar to those of the dark matter distribution in clusters of galaxies \cite{Dubinski1998}. We also know that the ICL follows the potential of the cluster. These two things together indicate that the ICL can be used as a tool to trace the global gravitational potential of its host cluster. 

Ref. \cite{Pillepich2014, Pillepich2018} proposed that the slope of the radial density profile of the ICL correlates with cluster mass, which was confirmed observationally in ref. \cite{MT18}. Moreover, ref. \cite{MT19} showed that the bi-dimensional distribution of ICL maps the distribution of the total mass in massive galaxy clusters up to 140 kpc from the centre of the cluster (within current observational uncertainties, $\sim25$ kpc). This result has been reproduced in simulations to an even larger radius (1 Mpc \cite{Alonso-Asensio2020}). What this is telling us is that by taking a ``picture'' of the ICL we can observe how the dark matter is distributed in the cluster. ICL maps can be used to discover dark matter substructures in clusters using deep images \cite{Jee2010}. Finding such substructures, how the mass distributes inside galaxy clusters, has important consequences in the study of dark matter itself \cite{Jauzac2016}.

Taking advantage of this property of the ICL, ref. \cite{Deason2021} used the C-EAGLE simulations to detect the splashback radius from the light profile of the clusters. The splashback radius is the radius where particles (stars or dark matter) reach the apocenter of their first orbit \cite{Diemer2014} separating material bound to the halo from infalling material; a physically motivated edge of a cluster. It also depends on the mass accretion rate because a rapidly growing potential well reduces the apocenters of the orbits. Therefore, it could be possible to infer the accretion history of the cluster by observing breaks in ICL radial profiles. Ref. \cite{Gonzalez2021} found a signature in the ICL profile at large radius ($\sim1.4$ Mpc) that could be interpreted as the splashback radius although at a smaller radius than expected from simulations. If confirmed, this will be the first identification of this radius for an individual cluster.

The ICL stands out as a promising way to infer, in great detail, the properties of the dark matter halos in galaxy clusters and the assembly history of clusters. Further studies, from both the simulations and observations point of view, will help us uncover the full potential of the ICL. 

\subsection*{To go beyond}

The aim of this review is to show the current state of the studies of the diffuse light in groups and clusters of galaxies. Evidently, we lack large samples to really understand in depth how the ICL forms, how it correlates with the properties of its host cluster and how it evolves with time. The recent advancements in technology and data processing have allowed us to extend the exploration of this light to intermediate \cite{Zibetti2005, MT14} and z$\sim$1 \cite{Burke2012, Adami2013, Ko2018} clusters and into the group regime \cite{daRocha2005, Poliakov2021}. However, studies still only analyse small samples (6-20, \cite{DeMaio2018, MT18, Furnell2021}) or employ stacking of many clusters to obtain coarse measurements \cite{Zibetti2005, Zhang2019, Sampaio-Santos2021}.

This scarcity of sufficiently good quality data will change with the next generation of surveys using state-of-the-art cameras that will allow us to reach unprecedented depths over large areas in the sky. They promise to deliver the large samples needed to explore the ICL with mass, redshift and dynamical state. For example, the Vera Rubin Observatory's LSST is expected to observe around 1 million galaxy groups up to redshift $z\sim1$ \cite{Brough2020}. That means that, for the first time, we will be able to undertake a statistically significant study of the properties of the ICL across mass ranges and redshifts. In addition, the combination of multiwavelength observations from LSST, Euclid and the Nancy Grace Roman Space Telescope will allow the most accurate determination of the stellar populations of the ICL. Having near-infrared observations of such faint systems mean being able to constrain the stellar populations in a way that is not possible with optical data alone. 

However, deep imaging is not enough. To study the ICL very accurate data processing is also needed. For example, the planned data reduction of Rubin's LSST data uses the LSST pipeline, which is currently being tested with HSC-SSP data. The sky subtraction algorithm in the HSC-SSP data release 1 severely over-subtracted extended halos of bright objects making it almost impossible to study nearby or extended objects \cite{Aihara2018} and significantly limited the study of stellar halos such that samples were limited to higher redshifts \cite{Huang2018} or \emph{a posteriori} techniques had to be used \cite{Furnell2021}. This issue was improved in the data release 2 \cite{Aihara2019} but not completely resolved (A.E. Watkins, priv. comm.). In addition, ICL studies are susceptible to biases due to flat-field inaccuracies. Ref. \cite{Montes2021} showed that dedicated processing pipelines can unveil the ICL to large radius ($\sim215$ kpc) with modest ($\sim30$ mins) exposure times. They show that custom flats and a careful sky background subtraction are needed to achieve a flat image, especially necessary if we aim to explore the ICL at even higher radial distances. In addition, a proper modelling of the telescopes point spread function is crucial to 1) correct the scattered light from bright stars and to 2) explore small/compact systems where the effect of the PSF can put extra light into the stellar halos and the ICL \cite{Sandin2014, Trujillo2016, Tang2018}.

Galatic cirri are an unavoidable problem in deep images. These cirri are the filamentary structures of dust from our Galaxy that reflect starlight. Their shape resembles that of extragalactic structures \cite{Duc2015} causing confusion. Moreover, these Cirri will become more and more pervasive in future ultra-deep surveys, even in regions far from the Galactic plane. Thankfully, ref. \cite{Roman2019} found that the optical colors of these filaments are bluer in $r-i$ than extragalactic sources, potentially leading to their identification at a higher spatial resolution than what is currently possible with far-infrared observations.

There are ongoing efforts both in Euclid \cite{Borlaff2021} and the Rubin's LSST (the LSB Working Group \cite{Brough2020}) to optimize these surveys for LSB science. Apart from data processing, these efforts will encounter other challenges that we are just starting to see. For example, the deblending of sources -- as we will be approaching the confusion limit from ground-based telescopes \cite{Borlaff2021} -- and the identification of LSB sources given the huge volumes of data expected. In addition, there are ongoing efforts to have an ultra-deep field with the Roman Space Telescope.

As we have seen in this review, a fundamental thing to address for ICL studies is the observational definition of the ICL. It is crucial to have a consistent approach when measuring ICL. This involves not only a consistent metric to separate BCG and ICL across redshifts and masses but also to be able to deal with different depths. The latter is important to compare with different surveys and, most importantly, at different redshifts. 

An interesting prospect is the detection of intracluster X-ray sources in Virgo and Fornax \cite{Hou2017, Jin2019}, likely associated with X-ray binaries in the ICL population \cite{Finoguenov2002}. This will not only open a new, and independent, window on the study of the ICL in other wavelengths rather than the optical or near-infrared, but could also provide constraints on progenitors of this light (their initial mass \cite{Peacock2014}). Unfortunately, the current spatial resolution of X-rays observatories limits these studies to nearby groups and clusters.

\vspace{3mm}
The ICL has been shown to be an unexpected tool to study groups and clusters of galaxy, allowing us to infer their assembly history, radius and even their dark matter distribution. It is also a prime example of what simulations and observations can achieve when working hand in hand. And still, we are only scratching the (bright) surface of what we can learn from this light.

\section*{Acknowledgements}
I would like to thank the referees (Zhiyuan Li, Enrichetta Iodice and Chris Mihos) for comments that helped improved the original manuscript. I would also like to thank Matthias Kluge for kindly providing the ICL fractions in \cite{Kluge2021}, Lin Tang for their ICL fraction predictions in \cite{Tang2018}, Takahiro Morishita for the ICL map of Abell S1063 in \cite{Morishita2017} and Ignacio Trujillo for useful discussions. 

\bibliographystyle{unsrt}
\bibliography{references}
\end{document}